\documentclass[aps,prc,reprint]{revtex4-1}
\usepackage{amssymb,amsbsy,amsfonts,amsmath}
\usepackage{epsfig,color}
\usepackage{epstopdf}
\usepackage[citecolor=blue,colorlinks=true,linkcolor=blue,urlcolor=blue]{hyperref}
\begin{document}
\title{Enhanced proton-boron nuclear fusion cross sections in intense high-frequency laser fields}
\author{Wenjuan Lv$^{1}$}
\author{Hao Duan$^{2,3}$}
\email{duan$_$hao@iapcm.ac.cn}
\author{Jie Liu$^{1}$}
\email{jliu@gscaep.ac.cn}
\affiliation{$^{1}$Graduate School, China Academy of Engineering Physics, Beijing 100193, China}
\affiliation{$^{2}$Laboratory of Computational Physics, Institute of Applied Physics and Computational Mathematics, Beijing 100088, China}
\affiliation{$^{3}$Institute of Applied Physics and Computational Mathematics, Beijing 100088, China}

\begin{abstract}
We investigate the proton-boron nuclear fusion cross sections under the influence of the intense linearly polarized monochromatic laser fields with high frequency. First, we rewrite the time-dependent Schr\"{o}dinger equation using Kramers-Henneberger (KH) transformation which allows for shifting all time dependence of the problem into the potential function. Then, for the intense laser fields that satisfy the high frequency limit, the time-averaged scheme in the KH framework should be valid. We can use WKB approximation to evaluate Coulomb barrier penetrability and then calculate proton-boron nuclear fusion cross sections by a phenomenological Gamow form.
We show that the corresponding Coulomb barrier penetrability increases significantly due to the depression of the time-averaged potential barrier. As a result, we find that proton-boron nuclear fusion cross sections can be enhanced effectively depending on a dimensionless quantity $n_{\mathrm{d}}$, which equals the ratio of the quiver oscillation amplitude to the geometrical touching radius of the proton and boron nucleus. For $n_{\mathrm{d}}=9$, we predict that the resonance peak of the fusion cross-section is enhanced by about $26$ times at the incident energy of $\varepsilon=148$ keV. And for another incident energy of $\varepsilon=586$ keV, the resonance peak of fusion cross-section is not only enhanced but also shifted to lower energy of $\varepsilon=392$ keV due to the mechanism of over-barrier fusion.
\end{abstract}
\maketitle
\section{Introduction}
With the rapid development of laser technology, intense lasers can be applied to atomic ionization \cite{Joachain, Liu, Mima}, charged-particle acceleration \cite{Mangles, Geddes, Faure}, and also provide a new way for manipulating nuclear processes, such as inducing resonance internal conversion \cite{Karpeshin} and nuclear nonlinear optics \cite{Tao Li JPG}, exciting the isomeric $^{229}$Th nuclear state based on laser-driven electron recollision process \cite{Wu Wang PRL, Wu Wang JPB}, accelerating nuclear fission processes \cite{Bai201801, Qi2020}, especially in the aspect of increasing $\alpha$-decay rates \cite{Delion2017, Bai201802, Qi2019} by modifying the Coulomb potential barrier.
However, there is no conclusive result on whether laser-induced enhancement of the decay rate is considerable or not because there is no successful experiment to prove these theories \cite{Palffy, Ghinescu}.

Since nuclear fusion processes are mainly associated with light nuclei, laser manipulation will be more effective because of the relatively large charge-mass ratio compared with that in the heavy nuclei processes.
%
%
For some of the main nuclear fusion reactions, the deuteron-triton (DT) reaction has been widely studied for its relatively high reaction cross sections compared to other fusion options \cite{Flowers, Argo, Bosch}.
%
%
Some theoretical investigations on the DT fusion process in the presence of intense laser fields have been done. Queisser and Sch$\ddot{\mathrm{u}}$tzhold reported that the tunneling probability could be enhanced in the x-ray free electron laser (XFEL) fields based on a Floquet scattering method \cite{Friedemann2019}. With the help of the Kramers-Henneberger (KH) transformation \cite{Kramers,Henneberger}, we found that DT fusion cross sections can be enhanced in the presence of electromagnetic fields with high-intensity and high-frequency \cite{Lv2019}.
Recently, it is shown that intense low-frequency laser fields, such as those in the near-infrared regime for the majority of intense laser facilities around the world, can also enhance the fusion probabilities \cite{Wang2020, Liushiwei2021, Qi2022}.

However, in DT fusion, tritium is a rare and radioactive isotope of hydrogen. One deuterium nucleus fuses with one tritium nucleus, yielding one helium nucleus, a free neutron, and the fusion power of about 17.6 MeV. Eighty percent of the fusion power (14.1 MeV) is associated with neutron, which is difficult to stop, making it difficult to avoid activation of the surrounding material. Activated material would cause significant radiation hazard and generate nuclear waste.
%
%
Therefore, the safety of fusion power systems would be greatly enhanced by the use of nonradiative and aneutronic fuel.

The proton-boron (p-$^{11}$B) nuclear reaction is particularly attractive because boron is both more plentiful and easier to handle than tritium. Natural boron is composed of two stable isotopes, $^{10}$B and $^{11}$B, the latter of which makes up about 80$\%$ of natural boron \cite{Lorenzo2020}.
Moreover, the fusion power is released mainly in charged $\alpha$ particles rather than neutrons \cite{Last2011, Labaune2013, Labaune2016, Hora2017}
\begin{equation}
\mathrm{p}+\mathrm{^{11}B} \rightarrow 3\alpha+8.6 \ \mathrm{MeV}.
\end{equation}
Although p-$^{11}$B nuclear fusion reaction is an environmentally clean reaction in contrast to DT fusion, its fusion cross section is several orders of magnitude smaller than DT fusion at the relatively modest centre-of-mass kinetic energy. The cross section of p-$^{11}$B nuclear reaction exhibits a very narrow resonance peak at incident energy $\varepsilon\approx148$ keV, and a broader resonance peak at $\varepsilon\approx586$ keV \cite{Atzeni}.
%

In this work, we investigate p-$^{11}$B nuclear fusion cross sections in the presence of high-frequency intense laser fields using the static KH approach.

The paper is organized as follows. Sec. II presents our model. Sec. III presents discussions on the Coulomb barrier penetrability. Our main results on p-$^{11}$B nuclear fusion cross sections are provided in Sec. IV. Sec. V presents our conclusion.
\section{Model}
In the presence of strong laser fields, the two-body spinless model Hamiltonian of p-$^{11}$B nuclear fusion in the Coulomb gauge is given by
\begin{eqnarray}
H&=&\frac{\left(\vec{p}_{1}-q_{1}\vec{A}\left(t_{1},\vec{r}_{1}\right)\right)^{2}}{2m_{1}}+\frac{\left(\vec{p}_{2}-q_{2}\vec{A}\left(t_{2},\vec{r}_{2}\right)\right)^{2}}{2m_{2}}\nonumber\\
&&+V\left(\vec{r}_{1}-\vec{r}_{2}\right),
\end{eqnarray}
where $m_{1(2)}$ and $q_{1(2)}$ are the nuclear masses and electrical charges of proton (boron nucleus) in the laboratory frame and $\vec{r}_{1(2)}$ and $\vec{p}_{1(2)}$ are the coordinate vectors and canonical momenta, respectively. $V\left(\vec{r}_{1}-\vec{r}_{2}\right)$ is the two-body interaction potential between the proton and boron nucleus, including the short-range attractive nuclear potential and long-range repulsive Coulomb potential, which can be given by
\begin{eqnarray}
V\left(\vec{r}_{1}-\vec{r}_{2}\right)&=&-\Theta\left(1-\frac{|\vec{r}_{1}-\vec{r}_{2}|}{r_{\mathrm{n}}}\right)U_{0}\nonumber\\
&&+\Theta\left(\frac{|\vec{r}_{1}-\vec{r}_{2}|}{r_{\mathrm{n}}}-1\right)\frac{q_{1}q_{2}}{4\pi\varepsilon_{0}|\vec{r}_{1}-\vec{r}_{2}|},
\end{eqnarray}
where the geometrical touching radius $r_{\mathrm{n}}=1.44\times(A^{1/3}_{1}+A^{1/3}_{2})\ \mathrm{fm}=1.44\times(1+11^{1/3})\ \mathrm{fm}\approx4.64\ \mathrm{fm}$ and $U_{0}\approx30\sim40$ MeV indicate the effective range and depth of the nuclear potential, respectively. $\Theta(x)$ is the unit step function.

The dipole approximation, which neglects spatial dependency of the laser electric field, can be used for the wave length of the laser field is greater than the typical size of the nucleus, so the vector potential depends only on time.
In the nonrelativistic limit, $t_{1}=t_{2}=t$.
The two-body Hamiltonian can be divided into a center part $H_{\mathrm{c}}=(\vec{P}-Q\vec{A}(t))^{2}/2M$ and a relative part $H_{\mathrm{r}}=(\vec{p}-q\vec{A}(t))^{2}/2m+V(\vec{r})$ with vanishing coupling between each other $[H_{\mathrm{c}},H_{\mathrm{r}}]=0$, where the relative displacement vector $\vec{r}$ of nuclei refers to $\vec{r}_{1}-\vec{r}_{2}$. The corresponding center motion and relative motion charges are $Q=q_{1}+q_{2}$ and $q=(q_{1}m_{2}-q_{2}m_{1})/(m_{1}+m_{2})$, and $M=m_{1}+m_{2}$ and $m=m_{1}m_{2}/(m_{1}+m_{2})$ are the total and reduced masses, respectively.

Therefore, the relative motion of the time-dependent Schr\"{o}dinger equation of p-$^{11}$B nuclear fusion in the center-of-mass frame is
\begin{eqnarray}
i\hbar\frac{\partial}{\partial t}\Psi\left(t,\vec{r}\right)=
\left(\frac{(\vec{p}-q\vec{A}\left(t\right))^{2}}{2m}+V\left(\vec{r}\right)\right)\Psi
\left(t,\vec{r}\right).\nonumber\\
\label{TDSE}
\end{eqnarray}
\subsection{KH transformation}
By adopting the unitary KH transformation
\begin{eqnarray}
\Omega\left(t\right)=\exp\left(\frac{i}{\hbar}\int^{t}_{-\infty}\left(-\frac{q}{m}\vec{A}\left(\tau\right)\cdot\vec{p}+\frac{q^{2}}{2m}\vec{A}^{2}\left(\tau\right)\right)d\tau\right),\nonumber\\
\end{eqnarray}
the wavefunction under the KH framework, denoted as $\Psi_{\mathrm{kh}}\left(t,\vec{r}_{\mathrm{kh}}\right)=\Omega(t)\Psi\left(t,\vec{r}\right)$, has the same total probability as $\Psi(t,\vec{r})$ due to $\Omega^{\dag}(t)\Omega(t)=1$. Then Eq. (\ref{TDSE}) becomes
\begin{eqnarray}
i\hbar\frac{\partial}{\partial t}\Psi_{\mathrm{kh}}\left(t,\vec{r}_{\mathrm{kh}}\right)=
\left(\frac{\vec{p}^{2}_{\mathrm{kh}}}{2m}+V_{\mathrm{kh}}
\left(t,\vec{r}_{\mathrm{kh}}\right)\right)\Psi_{\mathrm{kh}}
\left(t,\vec{r}_{\mathrm{kh}}\right),\nonumber\\
\label{TDSE-KH}
\end{eqnarray}

\begin{figure}[!b]
\centering
\includegraphics[width=0.8\columnwidth]{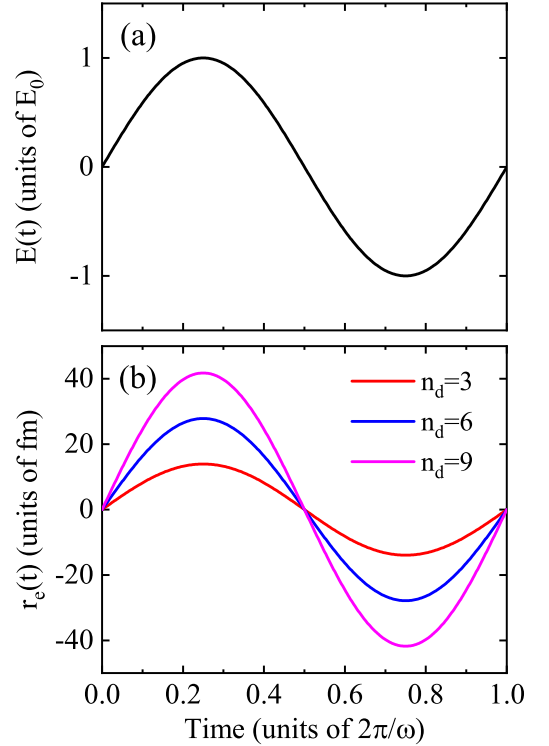}
\caption{(Color online) (a) The time-dependent electric fields (in units of $E_{0}$), (b) The quiver oscillations of p-$^{11}$B nuclear fusion system in the fields for $n_{\mathrm{d}}=3, 6$ and $9$, respectively.}
\label{fig1}
\end{figure}

where the time-dependent potential is found to be
\begin{eqnarray}
V_{\mathrm{kh}}\left(t,\vec{r}_{\mathrm{kh}}\right)&=&
-\Theta\left(1-\frac{|\vec{r}_{\mathrm{kh}}(t)|}
{r_{\mathrm{n}}}\right)U_{0}\nonumber\\
&&+\Theta\left(\frac{|\vec{r}_{\mathrm{kh}}(t)|}
{r_{\mathrm{n}}}-1\right)V_{0}\frac{r_{\mathrm{n}}}{|\vec{r}_{\mathrm{kh}}(t)|}
\label{Vkh}
\end{eqnarray}
with the height of the Coulomb barrier $V_{0}=5e^{2}/4\pi\varepsilon_{0}r_{\mathrm{n}}\approx1.55$ MeV.
Here, the momentum operator $\vec{p}_{\mathrm{kh}}=\vec{p}$,
and the coordinate operator $\vec{r}_{\mathrm{kh}}(t)=\vec{r}-\vec{r}_{\mathrm{e}}(t)$.
Eq. (\ref{TDSE-KH}) indicates that all of the time dependence is shifted into the potential function after KH transformation, and $V_{\mathrm{kh}}\left(t,\vec{r}_{\mathrm{kh}}\right)$ is just a two-body Coulomb potential dressed by a time-dependent harmonic oscillation origin $\vec{r}_{\mathrm{e}}(t)$ along the polarization direction $\hat{e}_{z}$ with a quiver oscillation amplitude $r_{\mathrm{e}}$.

Supposing that the laser field is monochromatic and linearly polarized along the $z$-axis
\begin{equation}
\vec{E}(t)=\hat{e}_{z}E_{0}\sin\omega t,
\label{Et}
\end{equation}
where $E_{0}$ is the amplitude and $\omega$ is the angular frequency. Then $\vec{r}_{\mathrm{e}}(t)=\hat{e}_{z}r_{\mathrm{e}}\sin\omega t$, where $r_{\mathrm{e}}=q\sqrt{2c\mu_{0}I}/(m\omega^{2})$ equals to the amplitude of quiver motion of a free nucleus in the laser fields. The commutation relation $[r^{i}_{\mathrm{kh}}(t),p^{j}_{\mathrm{kh}}(t)]=i\hbar\delta^{i,j}$ remains unchanged. Let us introduce a dimensionless quantity $n_{\mathrm{d}}=r_{\mathrm{e}}/r_{\mathrm{n}}=q\sqrt{2c\mu_{0}I}/(m\omega^{2}r_{\mathrm{n}})$,
where the units of $I$ and $\hbar\omega$ are $\mathrm{W/cm^{2}}$ and eV, respectively. The ratio of $r_{\mathrm{e}}$ to $r_{\mathrm{n}}$ determines how external fields manipulate nuclei fusion processes.

We show the shape of one cycle of the time-dependent monochromatic electric field and the quiver oscillations of p-$^{11}$B nuclear fusion system in the electric fields for $n_{\mathrm{d}}=3, 6$ and $9$ in Fig. \ref{fig1}. It can be seen from Fig. \ref{fig1} (a) that the electric field oscillates periodically in the range of $[-E_{0}, E_{0}]$. Fig. \ref{fig1} (b) shows the quiver oscillations for $n_{\mathrm{d}}=3, 6$ and $9$ accordingly, and the range are about $[-r_{\mathrm{e}}, r_{\mathrm{e}}]$.
\subsection{Time-Averaged Potential}
In the scattering process, there are two intrinsic time-scales: one is characterized by the laser frequency, the other is the Coulomb interaction time duration.

The characteristic Coulomb interaction time duration can be approximated by the ratio of interaction length to an average relative velocity, i.e., $\Delta t\approx10^{3}q_{\mathrm{1}}q_{\mathrm{2}}\sqrt{2m}/(4\pi\varepsilon_{\mathrm{0}}\varepsilon\sqrt{\varepsilon})$ \cite{Lv2022}, where $\varepsilon$ is the incident kinetic energy of the proton and boron nucleus.
The corresponding Coulomb interaction time durations are ranged from 31.61  femtoseconds to 1 attosecond for the incident kinetic energy $\varepsilon$ of proton and boron nucleus ranged from 1 keV to 1 MeV. For the laser frequency (photon energy) is larger than 10 keV, the corresponding field oscillating period is less than 0.41 attoseconds, which is fast comparing with the interaction time duration. We therefore believe that the time-averaged scheme in the KH framework should be valid, i.e., in the high-frequency laser fields, the incident nucleus feels a time-averaged potential $\overline{V_{\mathrm{kh}}\left(t,\vec{r}_{\mathrm{kh}}\right)}=V_{\mathrm{eff}}\left(\vec{r}\right).$

\begin{figure}[!tb]
\centering
\includegraphics[width=\columnwidth]{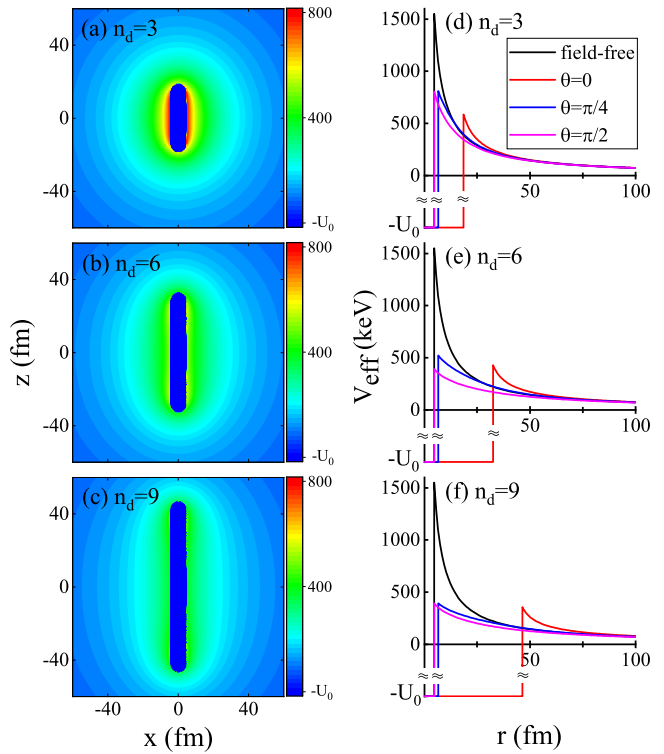}
\caption{(color online) The effective potential has rotational symmetry with respect to $z$ axis. (a) to (c) are contour plots on $x$-$z$ section ($y=0$) of  the effective potential for $n_{\mathrm{d}}=3, 6$ and $9$, respectively. The blue areas represent the section of inner region $\mathrm{D_{in}}$ where the potential value is approximately $-U_0$. $\mathrm{D_{in}}$ consists of a cylindrical region and two hemispheres at top and bottom ends. The length of the cylinder is $2r_{\mathrm{e}}$, and the radius of both the cylinder and the hemisphere is $r_{\mathrm{n}}$. (d) to (f) are $V_{\mathrm{eff}}$ for inclination angles $\theta=0,\pi/4,\pi/2$ with respect to varied $n_{\mathrm{d}}$.}
\label{fig2}
\end{figure}

In this case, the space can be divided into two parts: the inner region denoted by a capsule-like region swept by the nuclear potential well $-U_{0}$, i.e., $\mathrm{D_{in}}=\{\vec{r} |r\leq r_{\mathrm{in}}\}$, where $r_{\mathrm{in}}$ is the boundary of capsule-like region. And the outer region denoted by $\mathrm{D_{out}}=\mathrm{R}^{3}/\mathrm{D_{in}}$.

The choice of the laser parameters for $n_{\mathrm{d}}=3, 6$ and $9$ is similar to the Fig. 1 of our previous study \cite{Lv2019}.
When the laser frequency is $10$ keV, the corresponding laser intensities for $n_{\mathrm{d}}=3$, $6$ and $9$ are $5.00\times10^{26}\ \mathrm{W/cm^{2}}$, $2.00\times10^{27}\ \mathrm{W/cm^{2}}$ and $4.50\times10^{27}\ \mathrm{W/cm^{2}}$, respectively.
These intensities are higher than current XFEL fields, but considering the rapid development of laser, we should take a positive attitude towards the realization of higher intensity lasers.

The corresponding time-averaged potential $V_{\mathrm{eff}}$ for $n_{\mathrm{d}}=3, 6$ and $9$ are shown in Fig. \ref{fig2}, respectively.
The time-averaged potential in both its peak value and tunneling width is distorted in the presence of strong fields. Especially along the polarization direction $\hat{e}_{z}$, both the peak value and the barrier width are found to decrease significantly. And the angle dependence of time-averaged potential is symmetrical for $\theta=\pi/2$. Here, $\theta$ is the inclination angle, i.e., the angle between the relative motion direction of nuclei and the polarization direction of laser fields (+$z$ axis).
\section{Penetrability}
Using the WKB approximation \cite{Landau}, the penetrability of p-$^{11}$B nuclear fusion through the Coulomb barrier at the incident kinetic energy $\varepsilon$ along the direction $\hat{r}$ can be given by
\begin{eqnarray}
P\left(\theta;\varepsilon,n_{\mathrm{d}}\right)=\exp\left(-\frac{2}{\hbar}\int_{r_{\mathrm{in}}}^{r_{\mathrm{out}}}\sqrt{2m\left(V_{\mathrm{eff}}\left(\vec{r}\right)-\varepsilon\right)}dr\right),\nonumber\\
\label{P}
\end{eqnarray}
where $r_{\mathrm{in}}$ and $r_{\mathrm{out}}$ are the inner and outer turning points, respectively. Due to the symmetry of the Hamiltonian, the penetrability is independent of the azimuth $\varphi$. The penetrability explicitly depends on the inclination angle $\theta$, the incident kinetic energy $\varepsilon$ and the dimensionless parameter $n_{\mathrm{d}}$.

\begin{figure}[!tb]
\centering
\includegraphics[width=0.9\columnwidth]{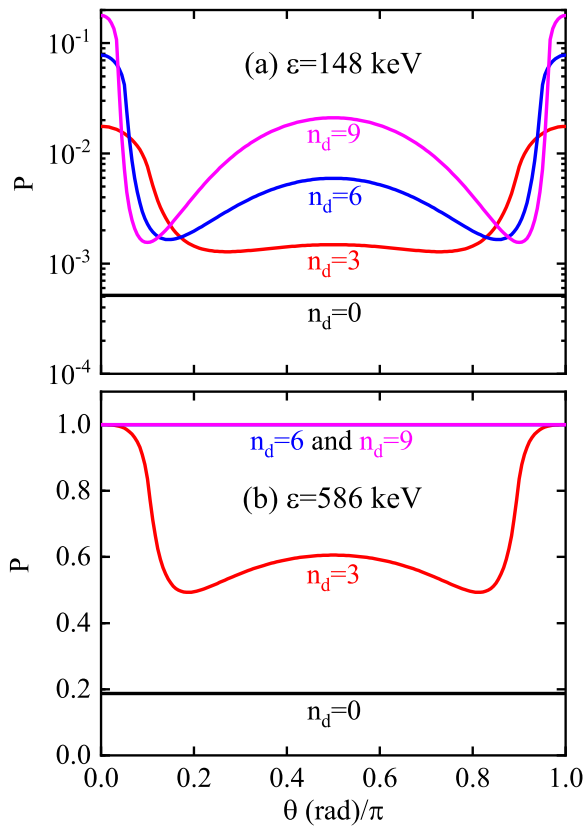}
\caption{(color online) Angle-dependent penetrabilities for incident kinetic energies of (a) $\varepsilon=148$ keV and (b) $\varepsilon=586$ keV. Notice that for $\varepsilon=586$ keV, the penetrabilities equal to 1 for all of the inclination angles when $n_{\mathrm{d}}=6$ and $n_{\mathrm{d}}=9$.}
\label{fig3}\
\end{figure}

The angle-dependent penetrability can be readily obtained by numerically calculating Eq. (\ref{P}), and the results of $n_{\mathrm{d}}=0,3,6$ and $9$ are plotted in Fig. \ref{fig3}. The angle-dependent penetrability for $\varepsilon=148$ keV exhibit an interesting double-hollow structure in Fig. \ref{fig3} (a). The penetrability can reach local maxima in the directions parallel and perpendicular to the laser polarization direction, i.e., $\theta=0,\pi$ and $\theta=\pi/2$. This is due to the distortion of the effective potential in both its peak and width in the presence of strong fields, as shown in Fig. 2.
The hollows correspond to the minimum penetrability, and their positions are approximately $\theta^{*}$ and $\pi-\theta^{*}$, with $\theta^{*}=2\arctan(r_\mathrm{n}/r_\mathrm{e})=2\arctan(1/n_\mathrm{d})$. This can be understood by observing the geometric character of the contour plots of the average potentials as shown in Fig. \ref{fig2} (a), (b) and (c). Fig. \ref{fig3} (b) shows that for $\varepsilon=586$ keV, the penetrability equals to 1 for all of the inclination angles when $n_{\mathrm{d}}=6$ and $n_{\mathrm{d}}=9$. That means the incident kinetic energy of proton and boron nucleus isn't less than the time-averaged potential for these cases. In fact, this over-barrier p-$^{11}$B nuclear fusion can be achieved when $n_{\mathrm{d}}\geq6$.

\begin{figure}[!tb]
\centering
\includegraphics[width=0.9\columnwidth]{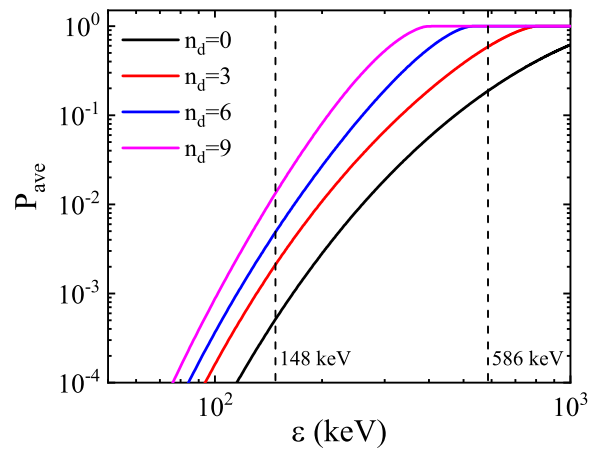}
\caption{(color online) Angle-averaged penetrability versus incident kinetic energy for $n_{\mathrm{d}}=0,3,6$ and $9$.}
\label{fig4}
\end{figure}

The angle average penetrability can be obtained by taking an average over the $4\pi$ solid angle, that is,
\begin{eqnarray}
P_{\mathrm{ave}}\left(\varepsilon;n_{\mathrm{d}}\right)&=&\frac{1}{2}\int_{0}^{\pi} P\left(\theta;\varepsilon,n_{\mathrm{d}}\right) \sin\theta d\theta.
\label{Pave}
\end{eqnarray}

The penetrability versus the relative kinetic energy for different $n_{\mathrm{d}}$ values are shown in Fig. \ref{fig4}, indicating that the penetrability increases significantly with respect to the dimensionless parameter $n_{\mathrm{d}}$.
For $n_{\mathrm{d}}=9$, the penetrability is approximately equals 0.01 for $\varepsilon=148$ keV, which is approximately $26$ times that of the field-free case. For $\varepsilon=586$ keV, $P_{\mathrm{ave}}=1$ when $n_{\mathrm{d}}\geq6$.
\section{Enhanced p-$^{11}$B nuclear fusion cross sections}
Nuclear fusion is commonly believed to consist of three processes. First, the wave packets of two nuclei collide with each other at a probability depicted by a geometrical cross section that depends on the de Broglie wavelength. Second, the approaching nucleus tunnels through the Coulomb potential barrier. Third, the nuclei come into contact and fuse, which can be described by an astrophysical $S$ factor.
Therefore, p-$^{11}$B nuclear fusion cross section is usually given in a phenomenological Gamow form \cite{Gamow} as a product of three terms:
\begin{eqnarray}
\sigma\left(\varepsilon\right)=\frac{S\left(\varepsilon\right)}{\varepsilon}P_{\mathrm{ave}}\left(\varepsilon;n_{\mathrm{d}}\right),
\label{Gamow}
\end{eqnarray}
where the term $1/\varepsilon$ is the geometrical cross section, which is proportional to the square of the relative motion's de Broglie wavelength. $S\left(\varepsilon\right)$ is the astrophysical $S$ factor that describes the nuclear physics within the nuclear potential effective range. In the absence of external laser fields, The $S$ factor can be given by the fitting function \cite{Nevins}.

In the presence of external fields, the astrophysical $S$ factor can still be described by the fitting function in field-free case. This is justified by the fact that what now regarded as very intense laser fields are still negligible compared to nuclear potentials. These laser fields have small effects on processes inside a nucleus or when the proton and boron nucleus are very close to each other.

\begin{figure}[!tb]
\centering
\includegraphics[width=0.9\columnwidth]{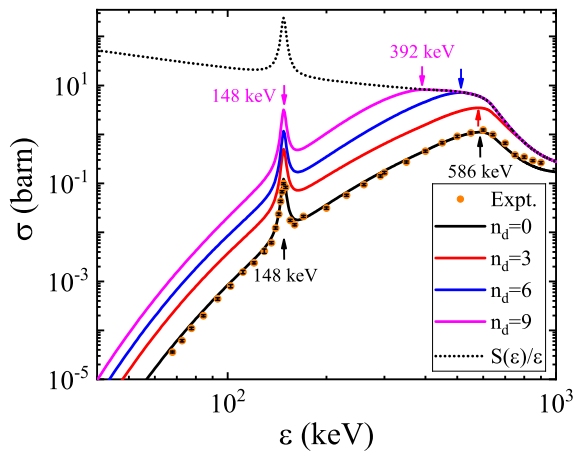}
\caption{(color online)p-$^{11}$B nuclear fusion cross sections versus incident kinetic energy for $n_{\mathrm{d}}=0,3,6$ and $9$. The experimental results and error bars are taken from Table 2 of Ref. \cite{Becker1987}. The black dotted line represents the achieve of over-barrier fusion. It's shown that the resonance peak of the fusion cross-section is enhanced but isn't shifted at the incident kinetic energy of $\varepsilon=148$ keV. But for another incident kinetic energy of $\varepsilon=586$ keV, it's not only enhanced but also shifted to lower energy due to the achieve of over-barrier fusion. The arrows in the figure represent the positions of the resonance peak.}
\label{fig5}
\end{figure}

The short-range nuclear potential is a complex square-well potential in the optical potential model \cite{Xinzhong Li PRC, Binbing Wu NPA}. The real and imaginary parts of potential represent particles scattering and absorption effects by potential, respectively. The lifetime of the particle inside the nuclear well is determined by the imaginary part of the potential $U_{i}$. Thus, $U_{i}/\hbar$ is chosen as the threshold of the laser frequency beyond which the time-averaged static potential is expected to be valid in investigating the change of the astrophysical $S$ factor in intense high-frequency laser fields \cite{Binbing Wu PRC}.

As shown in Fig. \ref{fig5}, the fusion cross sections have been significantly enhanced with the increasing of $n_{\mathrm{d}}$, which indicates that high-intensity and high-frequency laser fields can effectively increase the p-$^{11}$B nuclear fusion cross sections.
At the incident kinetic energy of $\varepsilon=148$ keV, the resonance peak of the fusion cross-section can be enhanced to $25.7$ barns, which is about $26$ times that of the field-free case for $n_{\mathrm{d}}=9$.
At the incident kinetic energy of $\varepsilon=586$ keV, the penetrability is approximately equals $1$ for $n_{\mathrm{d}}=9$, the fusion cross sections can be enhanced to $5.93$ barns, which is approximately $5$ times that of the field-free case. Moreover, the resonance peak is not only enhanced to $8.24$ barns in this laser field but also shifted from $\varepsilon=586$ keV to lower energy $\varepsilon=392$ keV due to the achieve of over-barrier fusion. This new resonance peak of the fusion cross-section is about $19$ times that of the field-free case.
\section{Conclusion}
In conclusion, we show that proton-boron nuclear fusion cross sections can be enhanced depending on a dimensionless parameter $\mathrm{n_{d}}$ in intense high-frequency laser fields.
For $n_{\mathrm{d}}=9$, one of the resonance peaks of the fusion cross-section is approximately $26$ times that of the field-free case for $\varepsilon=148$ keV. And another resonance peak appears at $\varepsilon=392$ keV rather than $\varepsilon=586$ keV due to the achieve of over-barrier fusion. This means that laser assisted p-$^{11}$B nuclear fusion can obtain bigger cross sections at smaller relative energies in such intense laser fields.

The hypothesis and limitation of our model are the high-frequency laser fields.
In reality, typically XFEL pulses are not fully temporally coherent and have a spiky structure. This issue is not an easy task and will be discussed in detail in our future work. On the other hand, extending these discussions to the situation of relatively low-frequency is undergoing.



%
%



\section*{Acknowledgments}
This work was supported by funding from NSAF No. U1930403.


\begin{thebibliography}{99}
\bibitem{Joachain}C. J. Joachain, N. J. Kylstra, and R. M. Potvliege, \textit{Atoms in Intense Laser Fields} (Cambridge University Press, Cambridge, 2012).
\bibitem{Liu}J. Liu, \textit{Classical Trajectory Perspective of Atomic Ionization in Strong Laser Fields} (Springer, Berlin, Heidelberg, 2014).
\bibitem{Mima}K. Mima, J. Fuchs, T. Taguchi, et al., Matter and Radiation at Extremes \textbf{3}, 127 (2018).
\bibitem{Mangles}S. P. D. Mangles, C. D. Murphy, Z. Najmudin, et al., Nature \textbf{431}, 535 (2004).
\bibitem{Geddes}C. G. R. Geddes, Cs. Toth, J. van Tilborg, et al., Nature \textbf{431}, 538 (2004).
\bibitem{Faure}J. Faure, Y. Glinec, A. Pukhov, et al., Nature \textbf{431}, 541 (2004).
\bibitem{Karpeshin}F. F. Karpeshin, Phys. Part. Nucl. \textbf{37}, 284 (2006).
\bibitem{Tao Li JPG}T. Li and X. Wang, J. Phys. G: Nucl. Part. Phys. \textbf{48} 095105 (2021).
\bibitem{Wu Wang PRL}W. Wang, J. Zhou, B. Q. Liu, and X. Wang, Phys. Rev. Lett. \textbf{127}, 052501 (2021).
\bibitem{Wu Wang JPB}W. Wang, H. X. Zhang and X. Wang, J. Phys. B: At. Mol. Opt. Phys. \textbf{54}, 244001 (2021).
\bibitem{Bai201801}D. Bai, D. M. Deng, Z. Z. Ren, Nuclear Physics A \textbf{976}, 23 (2018).
\bibitem{Qi2020}J. T. Qi, L. B. Fu, and X. Wang, Phys. Rev. C \textbf{102}, 064629 (2020).
\bibitem{Delion2017}D. S. Delion and S. A. Ghinescu, Phys. Rev. Lett. \textbf{119}, 202501 (2017).
\bibitem{Bai201802}D. Bai and Z. Z. Ren, Commun. Theor. Phys. \textbf{70}, 559 (2018).
\bibitem{Qi2019}J. T. Qi, T. Li, R. H. Xu, L. B. Fu, and X. Wang, Phys. Rev. C \textbf{99}, 044610 (2019).
\bibitem{Palffy}A. P$\acute{\mathrm{a}}$lffy and S. V. Popruzhenko, Phys. Rev. Lett. \textbf{124}, 212505 (2020).
\bibitem{Ghinescu}S. A. Ghinescu and D. S. Delion, Phys. Rev. C \textbf{101}, 044304 (2020).
\bibitem{Flowers}B. H. Flowers, Proc. R. Soc. Lond. A \textbf{204}, 503 (1951).
\bibitem{Argo}H. V. Argo, R. F. Taschek, H. M. Agnew, A. Hemmendinger and W. T. Leland, Phys. Rev. \textbf{87}, 612 (1952).
\bibitem{Bosch}H. S. Bosch and G. M. Hale, Nucl. Fusion \textbf{32}, 611 (1992).
\bibitem{Friedemann2019}F. Queisser and R. Sch$\ddot{\mathrm{u}}$tzhold, Phys. Rev. C \textbf{100}, 041601(R) (2019).
\bibitem{Kramers}H. A. Kramers, \textit{Collected Scientific Papers} (North Holland Publishing Company, Amsterdam, 1956).
\bibitem{Henneberger}W. C. Henneberger, Phys. Rev. Lett. \textbf{21} 838 (1968).
\bibitem{Lv2019}W. J. Lv, H. Duan, and J. Liu, Phys. Rev. C \textbf{100}, 064610 (2019).
\bibitem{Wang2020}X. Wang, Phys. Rev. C \textbf{102} 011601(R) (2020).
\bibitem{Liushiwei2021}S. W. Liu, H. Duan, D. F. Ye, and J. Liu, Phys. Rev. C \textbf{104}, 044614 (2021).
\bibitem{Qi2022}J. T. Qi, Nuclear Physics A \textbf{1020}, 122394 (2022).
\bibitem{Lorenzo2020}L. Giuffrida, F. Belloni, D. Margarone, et al., Phys. Rev. E \textbf{101}, 013204 (2020).
\bibitem{Last2011}I. Last, S. Ron, J. Jortner, Phys. Rev. A \textbf{83}, 043202 (2011).
\bibitem{Labaune2013}C. Labaune, C. Baccou, S. Depierreux, et al., Nature Communications \textbf{4}, 2506 (2013).
\bibitem{Labaune2016}C. Labaune, C. Baccou, V. Yahia, et al., Scientific Reports \textbf{6}, 21202 (2016).
\bibitem{Hora2017}H. Hora, S. Eliezer, G. J. Kirchhoff, et al., Laser and Particle Beams, \textbf{35}, 730 (2017).
\bibitem{Atzeni}S. Atzeni and J. Meyer-ter-Vehn, \textit{The physics of inertial fusion: Beam Plasma Interaction, Hydrodynamics, Hot Dense Matter} (Clarendon Press, Oxford, 2004).
\bibitem{Lv2022}W. J. Lv, B. B. Wu, H. Duan, S. W. Liu, and Jie Liu, Eur. Phys. J. A \textbf{58}, 54 (2022).
\bibitem{Landau}L. D. Landau and E. M. Lifshitz, \textit{Quantum Mechanics Non-relativistic Theory} (Pergamon Press, London-Paris, 1958).
\bibitem{Gamow}G. Gamow, Eur. Phys. J. A \textbf{51}, 204 (1928).
\bibitem{Nevins}W. M. Nevins and R. Swain, Nucl. Fusion \textbf{40} 865 (2000).
\bibitem{Xinzhong Li PRC}X. Z. Li, J. Tian, M. Y. Mei, and C. X. Li, Phys. Rev. C. \textbf{61}, 024610 (2000).
\bibitem{Binbing Wu NPA}B. B. Wu, H. Duan, and J. Liu, Nucl. Phys. A \textbf{1017}, 122340 (2022).
\bibitem{Binbing Wu PRC}B. B. Wu, H. Duan, and J. Liu, arXiv:2112.12384 (2021).
\bibitem{Becker1987}H. W. Becker, C. Rolfs, H. P. Trautvetter, Z. Phys. A Atoms Nucl. \textbf{327}, 341 (1987). Data retrieved from the US National Nuclear Data Center.
\end{thebibliography}
\end{document}